\documentclass[preprint,aps,prd,superscriptaddress,nofootinbib,tightenlines]{revtex4}
\usepackage[english]{babel}

\usepackage[letterpaper,top=2cm,bottom=2cm,left=3cm,right=3cm,marginparwidth=1.75cm]{geometry}

\usepackage{amsfonts}
\usepackage{multirow}
\usepackage{makecell}
\usepackage{mathrsfs}
\usepackage{graphicx}
\usepackage{amsmath}
\usepackage{amssymb}
\usepackage{bm}
\usepackage{bbm}
\usepackage{color}
\usepackage{ulem}
\usepackage{array}
\usepackage{diagbox}
\usepackage{float}

\newcommand{\nn}{\nonumber}

\newcommand{\beq}{\begin{equation}}
\newcommand{\eeq}{\end{equation}}
\newcommand{\bqa}{\begin{eqnarray}}
\newcommand{\eqa}{\end{eqnarray}}

\newcommand{\bseq}{\begin{subequations}}
\newcommand{\eseq}{\end{subequations}}

\makeatletter

\begin{document}

\title{Mixed Electroweak–QCD Corrections to $H\to \gamma\gamma$
}
\author{Wen-Long Sang~\footnote{wlsang@swu.edu.cn}}
\affiliation{School of Physical Science and Technology, Southwest University, Chongqing 400700, China\vspace{0.2cm}}
\author{Feng Feng~\footnote{f.feng@outlook.com}}
\affiliation{China University of Mining and Technology, Beijing 100083, China\vspace{0.2cm}}
\affiliation{Institute of High Energy Physics,
Chinese Academy of Sciences, Beijing 100049, China\vspace{0.2cm}}
\author{Yu Jia~\footnote{jiay@ihep.ac.cn}}
\affiliation{Institute of High Energy Physics, Chinese Academy of Sciences, Beijing 100049, China\vspace{0.2cm}}

\date{\today} 

\begin{abstract}
We present for the first time the complete three-loop mixed electroweak–QCD ($\mathcal{O}(\alpha\alpha_s)$) corrections 
for the decay channel $H \to \gamma\gamma$, by implementing three
different on-shell $\alpha$ schemes in computing the electroweak correction. Our studies indicate that the $\mathcal{O}(\alpha_s)$ correction amounts to approximately $1.7\%$ of the leading-order prediction for the diphoton width, while the $\mathcal{O}(\alpha)$ correction varies from $-4.8\%$ to $1.4\%$ 
depending on the specific $\alpha$ scheme. 
The three-loop mixed electroweak-QCD correction may reach $0.6\%$, $0.5\%$, and $0.2\%$ of the LO diphoton width in $\alpha(0)$, $\alpha(M_Z)$, and $G_\mu$ schemes, respectively, which is much more significant than the less-than-$0.1\%$ contribution from the three-loop QCD correction.  
It is also worth noting that the inclusion of the ${\cal O}(\alpha\alpha_s)$ correction significantly reduces the scheme dependence of the partial width from $0.6$ keV at leading order down to $0.03$ keV. The state-of-the-art Standard Model predictions are $\Gamma[H \to \gamma\gamma] = 9.389\div 9.420$ keV,  
providing a valuable theoretical benchmark for future Higgs factory collider program.  
\end{abstract}

\maketitle

\section{Introduction}

The ground-breaking discovery of the Higgs boson by {\tt ATLAS} and {\tt CMS} Collaborations in 2012~\cite{ATLAS:2012yve,CMS:2012qbp}
has heralded a new epoch of fundamental physics.
A high-priority theme of contemporary high energy physics is to measure all Higgs production and decay modes as precise as possible and seek the footprint of beyond Standard Model (SM) physics.
Notwithstanding a branching fraction merely of order $10^{-3}$, 
the Higgs boson decay to diphoton is among the original discovering channels of the Higgs boson, thanks to the very clean signal events and the  
exceptional precision of photon measurements at Large Hadron Collider (LHC).
Unexaggeratedly speaking, this gold-plated Higgs decay channel will continue to play a decisive role in testing SM at LHC and future colliders.

Historically, the theoretical investigations on $H\to\gamma\gamma$ decay within SM has undergone a long route. 
Being a loop-induced process, the leading-order (LO) amplitude for $H\rightarrow\gamma\gamma$ has been known since mid-1970s~\cite{Ellis:1975ap,Shifman:1979eb,Okun:1982ap,Gavela:1981ri}. 
The next-to-leading-order (NLO) (two-loop) QCD correction was calculated since early 90s~\cite{Zheng:1990qa,Djouadi:1990aj,Dawson:1992cy,Melnikov:1993tj,Inoue:1994jq,Spira:1995rr,Fleischer:2004vb,Harlander:2005rq,Aglietti:2006tp,Passarino:2007fp}.
Meanwhile, the NLO electroweak correction was first investigated in~\cite{Korner:1995xd,Liao:1996td,Djouadi:1997rj,Fugel:2004ug,Aglietti:2004nj,Aglietti:2004ki}, with the complete calculation achieved later in Refs.~\cite{Degrassi:2005mc,Passarino:2007fp}.
The next-to-next-to-leading order (NNLO) (three-loop) QCD correction was computed first with large top mass expansion in late 90s~\cite{Steinhauser:1996wy}, 
and was recently determined with full top mass dependence~\cite{Maierhofer:2012vv,Niggetiedt:2020sbf}. 
The $\rm N^3LO$ (four-loop) QCD correction was first partially inferred by Sturm {\it et al.} about a decade ago~\cite{Sturm:2014nva},  
and completed very recently with large top mass expansion~\cite{Davies:2021zbx}.  Partial $\rm N^4LO$ QCD contributions where both photons couple to a massive top loop can also be found in Ref.~\cite{Sturm:2014nva}. 
Four-loop analyses based on the principle of maximum conformality appear in Refs.~\cite{Yu:2018hgw,Luo:2023cpa}.
All-order QCD corrections in the large-$\beta_0$ limit were derived in Ref.~\cite{Boito:2022fmn}. 

The next generation of Higgs factories, {\it viz.}, the circular $e^+e^-$ collider program exemplified by {\tt CEPC} and {\tt FCC} experiments, 
aim to measure the branching fraction of $H\to\gamma\gamma$ to a percent or even sub-percent level~\cite{CEPC-SPPCStudyGroup:2015csa,CEPC-SPPCStudyGroup:2015esa,FCC:2018vvp}. 
Therefore, it appears imperative for theoretical predictions to the diphoton width to match  this projected experimental accuracy. 
It has long been known that the NLO electroweak and QCD corrections amount to several per cents of the LO contribution. 
One may naturally speculate that the mixed electroweak–QCD (three-loop) correction may reach the sub-percent level, 
therefore of great phenomenological incentive to add this missing piece of knowledge in order to meet the projected experimental precision. 
It is the central goal of this work is to evaluate the ${\cal O}(\alpha \alpha_s)$ mixed electroweak-QCD correction. 
For the sake of comparison, we also re-investigate the complete three-loop ${\cal O}(\alpha_s^2)$ correction.

The rest of the paper is distributed as follows.  
In Section~\ref{sec:theory:framework} we decompose the $H\to\gamma\gamma$ amplitude according to 
Lorentz structure and expresses the decay width in terms of a single scalar form factor.  
In Section~\ref{sec:LO:ampl} we recapitulate the LO result for $\Gamma(H\to\gamma\gamma)$,
emphasizing that the charged fermions like $b$, $c$ and $\tau$ should be retained.
We dedicate Section~\ref{sec:renorm} to recapitulate the on-shell renormalization scheme used in computing ${\cal O}(\alpha\alpha_s)$ correction, 
paying particular attention to the difference among three $\alpha$-subschemes.
In Section~\ref{sec:techniques} we briefly outline the technicality encountered in the calculation. 
In Section~\ref{sec:phenom} we present the predict the Higgs diphoton width at various level of perturbative accuracy, 
as well as report the state-of-the-art predictions for the $H\to\gamma\gamma$ decay rate and branching fraction.
Finally we summarize in Section~\ref{sec:summary}.

\section{Building diphoton width out of form factors~\label{sec:theory:framework}}

We start with the amplitude for $H\to \gamma\gamma$:
\beq
\mathcal{A}=T_{\mu\nu}\varepsilon^{\mu*}_\gamma(p_1) \varepsilon^{\nu*}_\gamma(p_2),
\eeq
where $p_{1,2}$ and $\varepsilon_\gamma(p_{1,2})$ signify momenta and polarization vectors
affiliated with two outgoing photons.
Lorentz invariance dictates the tensor $T^{\mu\nu}$ to be decomposed into the following most general form:
\bqa
 T^{\mu\nu} &=& p_1^\mu p_1^\nu T_1+p_2^\mu p_2^\nu T_2+p_1^\mu p_2^\nu T_3+p_2^\mu p_1^\nu T_4
\nn\\
&+& p_1\cdot p_2 g^{\mu\nu} T_5 +
\epsilon^{\mu\nu\alpha\beta}p_{1\alpha}p_{2\beta}T_6,
\label{eq:lorentz:structure}
\eqa
where all the nontrivial dynamics are encoded in the six scalar form factors $T_i$ ($i=1,\cdots 6$).

By sending outgoing photons on-shell, transversity condition implies that the $T_{1,2,3}$ terms make null contributions to the amplitude. 
Ward identity $p_{1\mu}T^{\mu\nu}=p_{2\nu}T^{\mu\nu}=0$ enforces $T_4=-T_5$. 
The $T_6$ term first arises from the two-loop electroweak correction in SM, which brings into the $CP$ violating effect.
Fortunately, its interference with both the LO and the $\mathcal{O}(\alpha_s)$ amplitude yields a vanishing contribution. 
Given the level of accuracy considered in this work, we can safely neglect $T_6$ from the outset. 
As a consequence,  the amplitude depends solely on a single form factor, {\it viz.}, $ T^{\mu\nu}=(p_2^\mu p_1^\nu -  p_1\cdot p_2 g^{\mu\nu}) T_4$.
The diphoton width of Higgs boson can then be expressed as
\beq
\Gamma(H\to \gamma\gamma)=\frac{1}{2}\frac{1}{2M_H}\frac{1}{8\pi}\frac{M_H^4}{2}|T_4|^2.
\label{Partial:width:from:T5:form:factor}
\eeq

In practice, it is advantageous to employ the following covariant projectors in $d$-dimensional spacetime to exact the scalar form factors 
$T_4$ and $T_5$~\cite{Steinhauser:1998rq}:  
\begin{subequations}
\bqa
{\tt P}_4^{\mu\nu}&=&\frac{2}{(d-2)M_H^2}\bigg[-g^{\mu\nu}+\frac{(d-1)p_1^\mu p_2^\nu}{p_1\cdot p_2}+\frac{p_2^\mu p_1^\nu}{p_1\cdot p_2}\bigg],
\\
{\tt P}_5^{\mu\nu}&=&\frac{2}{(d-2)M_H^2}\bigg[g^{\mu\nu}-\frac{p_1^\mu p_2^\nu}{p_1\cdot p_2}-\frac{p_2^\mu p_1^\nu}{p_1\cdot p_2}\bigg],
\eqa
\label{eq:projectors}
\end{subequations}
through
\bqa
T_4= {\tt P}_4^{\mu\nu} T_{\mu\nu},\quad\quad T_5= {\tt P}_5^{\mu\nu}T_{\mu\nu}.
\label{eq:lorentz:structure}
\eqa

\vspace{0.2 cm}

\section{LO result for $H\to\gamma\gamma$~\label{sec:LO:ampl}}

At LO, the Higgs diphoton decay is mediated via the one-loop diagrams with 
the $W$ boson and heavy fermions circuit around. 
It is customary to break the $T_4$ into two pieces:
\beq
T_4=\frac{e^3}{16\pi^2}\frac{1}{M_W s_W} (A_W+A_f),
\label{from:factor:T4:LO}
\eeq
where the individual contributions from the $W$ boson and fermion have long 
become the standard textbook knowledge: 
\begin{subequations}
\bqa
A_W&=&2+\frac{3}{x_W}+\frac{3}{x_W} \left( 2-\frac{1}{x_W}\right) \arcsin^2\sqrt{x_W},
\\
A_f&=&\sum_{f=t, b, c, \tau} 2 Q_f^2 C_f \bigg[\frac{1}{x_f}+\frac{1}{x_f}\left(1-\frac{1}{x_f}\right)\arcsin^2\sqrt{x_f}\bigg],
\label{from:factor:T4:LO}
\eqa
\end{subequations}
with the mass ratios defined by $x_W\equiv \tfrac{M_H^2}{4M_W^2}$ and $x_f\equiv \tfrac{M_H^2}{4M_f^2}$. 
$Q_f$ represents the electric charge of each fermion, while the color factor $C_f$ equals $3$ for quarks and  $1$ for leptons.
Note that in the sum in \eqref{from:factor:T4:LO}, we have explicitly included the contributions from heavy fermions 
such as bottom, charm-quark as well as $\tau$  lepton, apart from the top quark. The reason is that they yield contributions of 
the sub-percent level, which is compatible with the intended precision targeted in this work.

\vspace{0.2 cm}

\section{On-shell renormalization scheme~\label{sec:renorm}}

The renormalization procedure for computing the QCD corrections up to NNLO ($\mathcal{O}(\alpha_s^2)$),
is straightforward. 
The UV divergences can be eliminated by replacing the bare top quark mass with the pole mass through
$\mathcal{O}(\alpha_s^2)$ accuracy, accompanied by renormalizing  the strong coupling constant under the $\overline{\rm MS}$ scheme through $\mathcal{O}(\alpha_s^2)$
accuracy.

Electroweak (also mixed electroweak-QCD ) corrections are handled in the widely-used on-shell renormalization scheme~\cite{Ross:1973fp,Hollik:1988ii}, 
which fixes the renormalized parameters to be the precisely measured Higgs-boson mass , $W/Z$ masses, top quark mass as well as the QED coupling.
Specifically, we adopt the {\it Fleischer-Jegerlehner tadpole prescription}~\cite{Fleischer:1980ub,Dittmaier:2022maf,Denner:2019vbn} that,
instead of introducing tadpole counter-terms, 
we include all tadpole diagrams consistently in both the bare amplitudes and the self-energies used to evaluate the renormalization constants.

We employ three influential variants of electroweak on-shell scheme: $\alpha(0)$, $\alpha(M_Z)$, and $G_\mu$ scheme, 
which differ with each other in handling QED coupling renormalization.

In the $\alpha(0)$ scheme, the QED coupling is taken as the precisely measured fine structure constant in the Thomson limit, with the corresponding charge renormalization 
constant $\delta Z_e$ expressed as
\beq\label{eq:ze:orig}
\delta Z_e|_{\alpha(0)}= {1\over 2} \Pi^{AA}(0)-{s_W\over c_W}
{\Sigma^{A Z}_T(0) \over M_Z^2},
\eeq
where ${\Pi^{AA}(s)}\equiv \Sigma_T^{AA}(s)/s$ denotes the photon vacuum polarization. 
Since ${\Pi^{AA}(s)}$ is inevitably contaminated by non-perturbative hadronic contributions at low invariant mass, 
it is convenient to rewrite $\delta{Z_e}$ as
\bqa
\delta{Z_e}\big|_{\alpha(0)} &=& {1\over 2} \Delta{\alpha_{\rm had}^{(5)}}(M_Z)
+{1\over 2}{\rm Re}\, \Pi^{{AA}(5)}(M_Z^2)
\nn\\
&+&  {1\over 2}\Pi^{AA}_{\rm rem}(0)-{s_W\over c_W}
{\Sigma^{A Z}_T(0)\over M_Z^2},
\label{dZe:alpha(0):scheme}
\eqa
where $\Pi^{{AA}(5)}(M_Z^2)$ represents the photon vacuum polarization evaluated at momentum transfer
$M_Z^2$, by treating five flavors of quarks massless. 
$\Delta{\alpha_{\rm had}^{(5)}}(M_Z)$, extracted form the $R$-ratio measurement,  
encompasses the low-energy hadronic contribution.
$\Pi^{AA}_{\rm rem}(0)$  denotes the perturbbative contribution to vacuum polarization 
from $W$ boson, charged leptons and top quark.

The charge renormalization constants in $\alpha(M_Z)$ and $G_\mu$ schemes
can be readily converted from that in $\alpha(0)$ scheme: 
\bseq
\bqa
&& \delta{Z_e}\big|_{\alpha(M_Z)}=\delta{Z_e}\big|_{\alpha(0)}- {1\over 2} \Delta\alpha(M_Z),
\\
&& \delta Z_e|_{G_\mu}=\delta Z_e|_{\alpha(0)}-{1\over 2}\Delta r,
\label{Delta:alpha:split:into:two:terms}
\eqa
\eseq
where the $\mathcal{O}(\alpha)$ expressions of $\Delta\alpha(M_Z)$ and $\Delta r$ 
can be found in Ref.~\cite{Denner:1991kt}. 

The effective QED couplings in $\alpha(M_Z)$ and $G_\mu$ schemes are defined by
\begin{subequations}
\bqa
\label{dZe:alpha(MZ):scheme}
&& \alpha\left(M_Z\right)=\frac{\alpha(0)}{1-\Delta \alpha\left(M_Z\right)},
\\
\label{dZe:Gmu:scheme}
&& \alpha_{G_\mu}=\frac{\sqrt{2}}{\pi}G_{\mu} M_W^2\left(1-\frac{M_W^2}{M_Z^2}\right),
\eqa
\end{subequations}
where large (non-)logarithmic corrections from light fermion and top quark loops 
are resumed,  so that perturbative convergence behavior is improved.
For more details, we refer the interested readers to Refs.~\cite{Denner:1991kt,Sun:2016bel,Chen:2018xau}.

To investigate the mixed electroweak-QCD corrections to $H\to\gamma\gamma$, we need to extend the knowledge of field and mass renormalization constants to the 
$\mathcal{O}(\alpha\alpha_s)$ accuracy.  
Various counterterms, {\it e.g.}, $\delta Z_e$, $\delta Z_{AZ}$, $\delta Z_{AA}$, $\delta Z_H$, $\delta M_Z$, $\delta M_W$ can be  directly  read off 
from the analytic $\mathcal{O}(\alpha\alpha_s)$ gauge-boson and Higgs self energies given in Refs.~\cite{Djouadi:1993ss,Kniehl:1994ph,Borrill:1994nk}.  
To ensure consistent treatment of tadpole contributions,  following the procedure of \cite{Denner:1991kt} we rederive these renormalization 
constants together with the top mass counterterm $\delta m_t$ through $\mathcal{O}(\alpha\alpha_s)$ accuracy.
Finally, we take the $\mathcal{O}(\alpha\alpha_s)$ correction to $\Delta r$ from Ref.~\cite{Dittmaier:2014qza}.

\vspace{0.2 cm}

\section{Outline of Calculation~\label{sec:techniques}}


\begin{figure}[hbtp]
\centering
\includegraphics[width=0.8\textwidth]{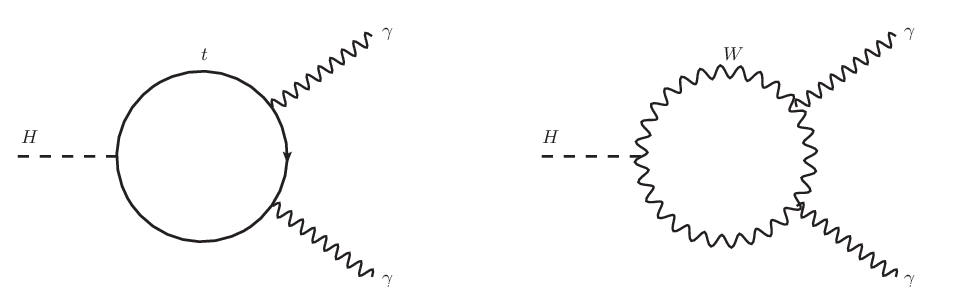}
\caption{Some representative LO diagrams for $H\to \gamma\gamma$.}
\label{Feynman:Diagram:LO}
\end{figure}

\begin{figure}[hbtp]
	\centering
	\includegraphics[width=1.0\textwidth]{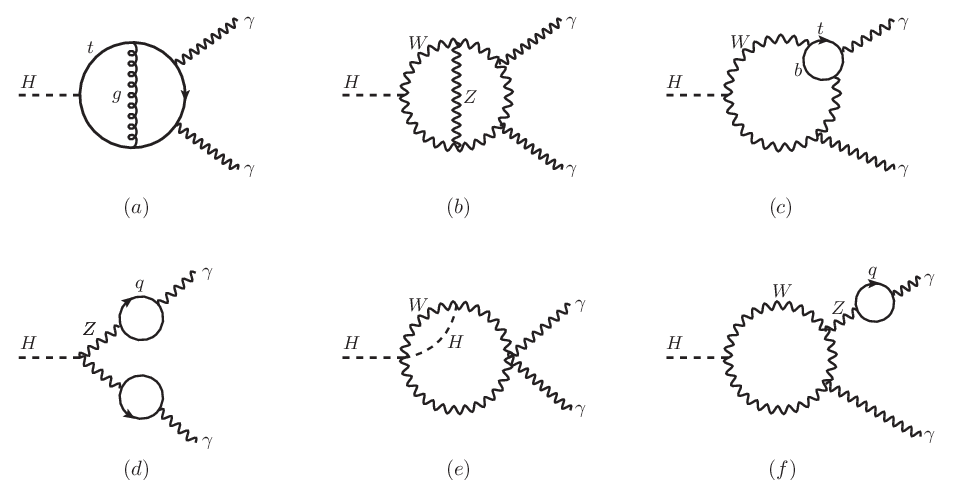}
	\caption{Some representative two-loop diagrams for $H\to \gamma\gamma$.
     $(a)$ represents a sample diagram for the NLO QCD correction, while $(b)$-$(f)$ represent sample
     diagrams for NLO electroweak correction.}
	\label{Feynman:Diagram:NLO}
\end{figure}

\begin{figure}[hbtp]
	\centering
	\includegraphics[width=1.0\textwidth]{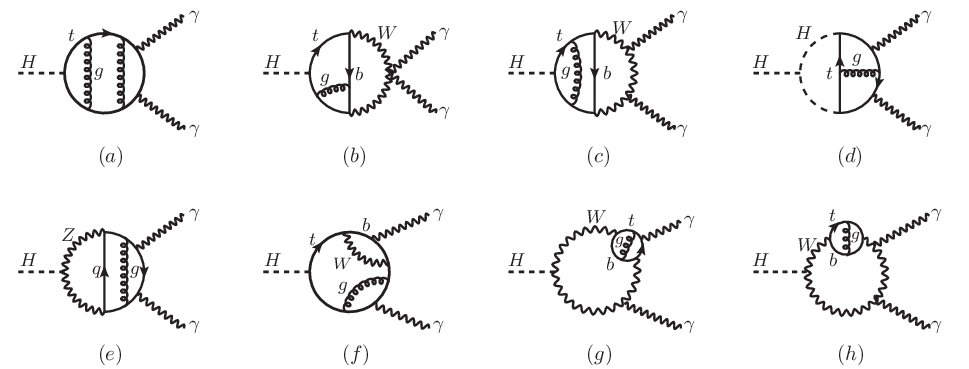}
	\caption{Some representative three-loop diagrams for $H\to \gamma\gamma$.
     $(a)$ represents a sample diagram for the NNLO QCD correction, while $(b)$-$(f)$ represent sample
     diagrams for the NNLO ${\cal O}(\alpha \alpha)_s$ correction.}
	\label{Feynman:Diagram:NNLO}
\end{figure}

Throughout this work we work with the Feynman gauge and regulate UV and IR divergences via dimensional regularisation in $d=4-2\epsilon$ dimensions.
Feynman diagrams and the corresponding amplitudes for $H\to \gamma\gamma$ through three loop are generated with {\tt FeynArts}~\cite{Hahn:2000kx}.
There are $28$ one-loop diagrams, $7305$ two-loop diagrams, and $6854$ three-loop diagrams ($326$ pure QCD diagrams and $6528$ mixed electroweak-QCD diagrams).
Representative LO diagrams for the LO process is depicted in Fig.~\ref{Feynman:Diagram:LO}. Some representative diagrams representing
NLO QCD and electrowaek corrections are shown in Fig.~\ref{Feynman:Diagram:NLO}, and those representing 
the NNLO QCD and mixed electroweak–QCD corrections are depicted in Fig.~\ref{Feynman:Diagram:NNLO}.

Lorentz algebra and Dirac traces are handled with the aid of the package {\tt FeynCalc}~\cite{Mertig:1990an} and {\tt FormLink}~\cite{Feng:2012tk}.
The form factors $T_{4,5}$ are extracted from the amplitude by employing the covariant projectors \eqref{eq:projectors}.
The package {\tt Apart}~\cite{Feng:2012iq} is used for partial fraction.  
All loop integrals are classified with {\tt CalcLoop}~\cite{CalcLoop}, and the integration-by-parts reduction is 
executed with the packages {\tt Kira}~\cite{Klappert:2020nbg}, {\tt Blade}~\cite{Guan:2024byi}, and {\tt FIRE}~\cite{Smirnov:2014hma}.
The resulting master integrals are evaluated to high numerical accuracy 
with {\tt AMFlow}~\cite{Liu:2017jxz,Liu:2022mfb,Liu:2022chg}.

After completing the renormalization procedure as outlined in Sec.~\ref{sec:renorm}, 
we end up with the UV and IR-finite results for $T_4$ and $T_5$.
A stringent test of the correctness of our calculation is provided by Ward identity, $T_4=-T_5$.
Our numerical study indicates that,  although the bare quantities appear to disrespect this relation for electroweak corrections, the Ward identity is exactly restored once the full set of counterterms is applied, 
to every perturbative order.

\vspace{0.2cm}

\section{Phenomenology~\label{sec:phenom}}

In this section we conduct a detailed numerical investigation on 
the diphoton width of Higgs at various perturbative accuracy,
paying special attention to the impact of the ${\cal O}(\alpha_s^2)$ and ${\cal O}(\alpha\alpha_s)$ 
corrections.

\subsection{Input parameters}

In our numerical study, we choose the following values for various masses~\cite{ParticleDataGroup:2022pth}:
\bqa
&&M_H=125.25\:{\rm GeV}, \quad M_Z=91.1876\: {\rm GeV}, \quad M_W=80.377\: {\rm GeV}, \quad m_t=172.69 \:{\rm GeV},\nn\\
&&m_e=0.5109989 \: {\rm MeV}, \quad m_{\mu}=105.65837 \: {\rm MeV},
\quad m_{\tau}=1.77686 \: {\rm GeV}.
\eqa
We further set $m_c=1.5$ GeV and $m_b=4.7$ GeV in the LO prediction.  
Charged lepton masses are retained in the computation of the charge renormalization constant $Z_e$ 
in the $\alpha(0)$ scheme. 
Apart from the situations of deducing the LO contribution and determination of $Z_e$, 
all quarks (except top quark) and charged leptons are treated as massless. 

We take the fine-structure constant in the Thomson limit, 
$\alpha(0)=1/137.035999$. With the aid of \eqref{dZe:Gmu:scheme}, 
the Fermi coupling $G_{\mu}=1.1663787\times 10^{-5}~{\rm GeV}^{-2}$ translates into 
$\alpha_{G_\mu}=1/132.168$.
The nonperturbative hadronic contribution to the photon 
vacuum polarization is chosen as $\Delta\alpha_{\rm had}^{(5)}(M_Z)=0.02764$~\cite{ParticleDataGroup:2022pth}.

In line with \eqref{dZe:alpha(MZ):scheme}, the prerequisite of determining
$\alpha(M_Z)$ is to know the concrete values of 
$\Delta \alpha_q(M_Z)$ and $\Delta \alpha_l(M_Z)$. 
The nonperturbative parameter $\Delta \alpha_q(M_Z)$ 
can be interchangeably used with the experimentally determined $\Delta\alpha_{\rm had}^{(5)}(M_Z)$. 
$\Delta \alpha_{\ell}(M_Z)$ can be calculated in perturbation theory, 
whose value is determined through four-loop QED accuracy~\cite{Kuhn:1998ze,Eidelman:1995ny,Steinhauser:1998rq,Sturm:2013uka}.
We adopt the four-loop result $\Delta \alpha_{\ell}\left(M_Z\right)\approx 0.0314979$~\cite{Sturm:2013uka}.
Consequently, by employing (\ref{dZe:alpha(MZ):scheme}), we obtain $\alpha(M_Z)=1/128.932$.

 When computing the QCD corrections, we freeze the renormalization scale at the Higgs mass, 
and the strong coupling constant is set to $\alpha_s(M_H)=0.115$~\cite{Bonciani:2015eua}. 
The uncertainty from sliding the renormalization scale is too tiny to
be considered here.

At present, the experimentally measured total width of the Higgs boson, 
$\Gamma_H=3.2^{+2.4}_{-1.7}$ MeV~\cite{ParticleDataGroup:2022pth}, 
is still subject to large uncertainties.  
To minimise theoretical uncertainty when predicting the branching ratio of $H\to \gamma\gamma$, 
we instead turn to much more precise theoretical prediction, $\Gamma_H=4.07^{+4.0\%}_{-3.9\%}$ MeV,
which was provided by the {\tt LHC} Higgs Working Group~\cite{ParticleDataGroup:2022pth,LHCHiggsCrossSectionWorkingGroup:2016ypw}.

\subsection{Numerical predictions for $T_4$}

Throughout this work we choose to freeze the QED coupling constants 
associated with two on-shell photon emission vertices to be $\alpha(0)$. However, within the electroweak on-shell renormalization scheme, 
there still exists some freedom in choosing some other sub-schemes 
to handle the charge renormalization in the third vertex.

In below we present the expressions of the form factor $T_4$ in three different $\alpha$ aub-schemes.
Firstly, $T_4$ in $\alpha(0)$ scheme reads
\bqa
\label{T4:alpha(0):scheme}
T_4 &=& - T_5 = -0.0488\,\alpha^{3/2}(0)\bigg[1+0.235\;\frac{\alpha_s}{\pi} 
+0.235\;\frac{\beta_0}{4}\frac{\alpha_s^2}{\pi^2}\ln\frac{\mu_R^2}{M_H^2} +0.122\frac{\alpha_s^2}{\pi^2} 
\nn\\
&-& 2.934 \;\frac{\alpha(0)}{\pi} + \frac{\Delta{\alpha_{\rm had}^{(5)}}(M_Z)}{2}
+33.280\frac{\alpha}{\pi}\frac{\alpha_s}{\pi}+0.118 \Delta{\alpha_{\rm had}^{(5)}}(M_Z)\frac{\alpha_s}{\pi}
\bigg],
\eqa
where $\beta_0$ signifies 
the one-loop coefficient of the QCD $\beta$ function with the number of active flavors $n_f=6$. 
Note that in Eq.~\eqref{T4:alpha(0):scheme}, the imaginary part has been omitted 
due to its negligible impact on the decay width. This treatment is justified 
since both the LO and $\mathcal{O}(\alpha_s)$ contributions 
from the $t$ quark and the $W$ boson,  are purely real. 
Although the contributions to LO prediction from lighter fermions, such as the $b$, $c$, and $\tau$, 
do generate an imaginary piece,  
the interference between them and the imaginary parts 
arising from the $\mathcal{O}(\alpha)$ and $\mathcal{O}(\alpha\alpha_s)$ 
corrections remains well below the sub-percent level, 
therefore can be safely neglected. 

Alternatively, $T_4$ in $\alpha(M_Z)$ scheme reads
\bqa
\label{T4:alpha(MZ):scheme}
T_4 &=&  -T_5 = -0.0488\,\alpha(0)\alpha^{1/2}(M_Z)\bigg[1+0.235\;\frac{\alpha_s}{\pi}
+0.235\;\frac{\beta_0}{4}\frac{\alpha_s^2}{\pi^2}\ln\frac{\mu_R^2}{M_H^2} +0.122\frac{\alpha_s^2}{\pi^2} 
\nn\\
&-& 
9.697 \;\frac{\alpha(M_Z)}{\pi}
  +31.690\frac{\alpha(M_Z)}{\pi}\frac{\alpha_s}{\pi}
\bigg].
\eqa

Lastly, $T_4$ in $G_\mu$ scheme reads
\bqa
\label{T4:gmu:scheme}
T_4 &= & -T_5 =  
-0.0488\,\alpha(0)\alpha^{1/2}_{G_\mu}\bigg[1+0.235\;\frac{\alpha_s}{\pi} +0.235\;\frac{\beta_0}{4}\frac{\alpha_s^2}{\pi^2}\ln\frac{\mu_R^2}{M_H^2} +0.122\frac{\alpha_s^2}{\pi^2}
 \nn\\
&- & 3.448 \;\frac{\alpha_{G_\mu}}{\pi}
+11.018\frac{\alpha_{G_\mu}}{\pi}\frac{\alpha_s}{\pi}
\bigg].
\eqa

\subsection{Numerical predictions for diphoton width and branching fraction}

\begin{widetext}
\begin{table*}[htbp]
\centering
\small
\setlength{\tabcolsep}{10pt}
\renewcommand{\arraystretch}{1.6}
\resizebox{\textwidth}{!}{%
\begin{tabular}{lcccccccc}
\hline
 &
$\Gamma^{\rm LO}$ &
$\Gamma^{\mathcal{O}(\alpha_s)}$ &
$\Gamma^{\mathcal{O}(\alpha)}$ &
$\Gamma^{\mathcal{O}(\alpha_s^2)}$ &
$\Gamma^{\mathcal{O}(\alpha\alpha_s)}$  & $\Gamma^{\rm Sum}$ & ${\cal B}(\times 10^{-3})$\\
\hline
$\alpha(0)$ scheme &
$9.060$ &
$0.156$ &
$0.127$ &
$0.004$ &
$0.055$&
$9.401$&
$2.31\pm 0.09$\\
\hline
$\alpha(M_Z)$ scheme &
$9.629$ &
$0.166$ &
$-0.461$ &
$0.004$ &
$0.051$&
$9.389$&
$2.31\pm 0.09$\\
\hline
$G_\mu$ scheme &
$9.393$ &
$0.162$ &
$-0.156$ &
$0.004$ &
$0.017$&
$9.420$&
$2.31\pm 0.09$\\
\hline
\end{tabular}%
}
\caption{Predicted partial width (in units of keV) for $H\to \gamma\gamma$ in various $\alpha$ schemes 
at various levels of perturbative accuracy.
$\Gamma^{\rm LO}$ represents the LO prediction.
The numerical values of the  NLO QCD corrections, NLO electroweak correction,
NNLO QCD corrections, together with the three-loop mixed electroweak-QCD corrections, 
are denoted by $\Gamma^{\mathcal{O}(\alpha)}$, $\Gamma^{\mathcal{O}(\alpha_s)}$, $\Gamma^{\mathcal{O}(\alpha\alpha_s)}$, and $\Gamma^{\mathcal{O}(\alpha_s^2)}$, respectively.
We have fixed the QCD renormalization scale $\mu_R=M_H$ and take $\alpha_s(M_H)=0.115$.
$\Gamma^{\rm Sum}$ is the most complete prediction by implementing all the aforementioned 
higher-order corrections. 
The branching fraction in the rightmost column is obtained via dividing $\Gamma^{\rm Sum}$ 
by the predicted full Higgs width, $\Gamma_H=4.07^{+4.0\%}_{-3.9\%}$ MeV~\cite{ParticleDataGroup:2022pth,LHCHiggsCrossSectionWorkingGroup:2016ypw}.
\label{table:decay-width}}
\end{table*}
\end{widetext}

Inserting $T_4$ in \eqref{T4:alpha(0):scheme},  \eqref{T4:alpha(MZ):scheme}
and \eqref{T4:gmu:scheme} into \eqref{Partial:width:from:T5:form:factor}, 
we then obtain the state-of-the-art predictions to the 
$\Gamma(H\to \gamma\gamma)$ decay rate in three $\alpha$ schemes.
In Table~\ref{table:decay-width} we enumerate 
the predicted decay width (in units of keV) at various perturbative accuracy

Firstly, we note that, despite strongly suppressed Yukawa couplings of the $b$, $c$ and $\tau$ relative to 
that of $t$, their individual contributions to the LO prediction amount to $0.8\%$, $0.7\%$, and $0.7\%$, respectively. Therefore, to the anticipated sub-percent accuracy, it is compulsory to include the lighter fermion
contribution at LO.

Secondly, it is interesting to point out that, 
the NLO electroweak correction and NLO QCD correction 
are of comparable magnitude~\footnote{This is in sharp contrast to the decay process $H\to Z\gamma$, 
where the NLO QCD correction constitutes only $0.3\%$ fraction of the LO prediction, 
and is much smaller than the NLO electroweak correction~\cite{Sang:2024vqk,Chen:2024vyn}.}. 
Moreover, the NLO electroweak corrections for $H\to\gamma\gamma$ 
from three different $\alpha$ schemes appear to differ significantly. In particular,
the NLO electroweak correction reaches approximately $-5\%$ in the $\alpha(M_Z)$ scheme, 
three times greater than in the other two schemes in magnitude.
A pronounced cancellation between the NLO electroweak and NLO QCD 
corrections occurs in the $\alpha(M_Z)$ scheme, 
while the cancellation is almost complete in the $G_\mu$ scheme.

Thirdly, we notice that the NNLO QCD ($\mathcal{O}(\alpha^2_s)$) correction is too small to 
bear phenomenological impact.
In contrast, the mixed electroweak-QCD ($\mathcal{O}(\alpha\alpha_s)$) 
correction can reach $0.5\%$ in the $\alpha(0)$ and $\alpha(M_Z)$ schemes, 
which is about an order of magnitude more important than the $\mathcal{O}(\alpha_s^2)$ correction, 
and remains sizeable, albeit somewhat smaller, in the $G_\mu$ scheme. 

Finally, we stress that including electroweak corrections 
plays a vital role in reducing the scheme dependence in electroweak sector.
The LO predictions span a $0.6$ keV range across three different $\alpha$ schemes.  
After including all the higher order corrections, 
the scheme dependence has been significantly reduced, with the residual spread less than $0.03$ keV.

Incorporating all the aforementioned higher order corrections, we are able to present 
the most complete and accurate SM prediction to $H\to \gamma\gamma$. 
The state-of-the-art diphoton width of Higgs is $9.389 \div 9.420$ keV, and 
the branching fraction is $(2.31\pm 0.09)\times 10^{-3}$, 
where the uncertainty is dominated by the theoretical uncertainty 
inherent in the predicted Higgs full width.

\section{summary~\label{sec:summary}}

In this work, we for the first time accomplish the calculation of the three-loop mixed electroweak–QCD corrections for the decay channel $H \to \gamma\gamma$. Three different electroweak on-shell renormalization
schemes are implemented. We also reinvesigate the known ${\cal O}(\alpha)$, 
${\cal O}(\alpha_s)$ and ${\cal O}(\alpha^2_s)$ corrections.

We find that the NLO electroweak correction ranges from $-4.8\%$ to $1.4\%$ of the LO prediction  
in three different $\alpha$ schemes, whereas the NLO QCD correction is about $+1.7\%$. 
The three-loop mixed electroweak–QCD correction exceeds $0.5\%$ of the LO prediction 
in the $\alpha(0)$ and $G_\mu$ schemes, and amounts to $0.18\%$ in the $G_\mu$ scheme. 
In contrast, the three-loop QCD correction only yields a contribution less than $0.1\%$, 
well below the phenomenological sensitivity. 

Including the mixed electrowek-QCD correction turns to be helpful to reduce the scheme dependence.
The predicted LO widths span $0.6$ keV across the three schemes. After including all corrections,
the residual scheme dependence is reduced to be less than 0.03 keV.
Piecing all corrections together, the state-of-the-art predictions are 
$\Gamma[H\to \gamma\gamma]=9.389\div 9.420$ keV, and ${\cal B}[H\to \gamma\gamma]= (2.31\pm 0.09)\times 10^{-3}$.
This new knowledge serves a useful reference for the prospective Higgs factories {\tt CEPC}
and {\tt FCC} to stringently test the Standard Model,
with the projected measurement precision reaching per-mille level.

\begin{acknowledgments}
Feynman diagrams in this work are drawn with the aid of {\tt JaxoDraw}~\cite{Binosi:2008ig}.
The work of W.-L. S. is supported by the NNSFC under Grant No.~12375079, and the
Natural Science Foundation of ChongQing under Grant No. CSTB2023 NSCQ-MSX0132.
The work of F.~F. is supported by the NNSFC under Grant No. 12275353.
The work of Y.~J. is supported in part by the NNSFC under Grant No.~12475090.
\end{acknowledgments}

\vspace{0.2 cm}


\end{document}